\documentclass[twocolumn,twocolappendix]{aastex631}

\usepackage{url}

\newcommand\revise{}

\begin{document}

\title{Early results from GLASS-JWST VIII: An Extremely Magnified Blue Supergiant Star at Redshift 2.65 in the Abell 2744 Cluster Field}

\correspondingauthor{Wenlei Chen}
\email{chen6339@umn.edu}

\author[0000-0003-1060-0723]{Wenlei Chen}
\affiliation{School of Physics and Astronomy, University of Minnesota, 116 Church Street SE, Minneapolis, MN 55455, USA}

\author[0000-0003-3142-997X]{Patrick L. Kelly}
\affiliation{School of Physics and Astronomy, University of Minnesota, 116 Church Street SE, Minneapolis, MN 55455, USA}

\author[0000-0002-8460-0390]{Tommaso Treu}
\affiliation{Department of Physics and Astronomy, University of California, Los Angeles, 430 Portola Plaza, Los Angeles, CA 90095, USA}

\author[0000-0002-9373-3865]{Xin Wang}
\affil{School of Astronomy and Space Science, University of Chinese Academy of Sciences (UCAS), Beijing 100049, China}
\affil{National Astronomical Observatories, Chinese Academy of Sciences, Beijing 100101, China}
\affil{Infrared Processing and Analysis Center, Caltech, 1200 E. California Blvd., Pasadena, CA 91125, USA}

\author[0000-0002-4140-1367]{Guido Roberts-Borsani}
\affiliation{Department of Physics and Astronomy, University of California, Los Angeles, 430 Portola Plaza, Los Angeles, CA 90095, USA}

\author[0000-0002-9800-9868]{Allison Keen}
\affiliation{School of Physics and Astronomy, University of Minnesota, 116 Church Street SE, Minneapolis, MN 55455, USA}

\author[0000-0001-8156-6281]{Rogier A. Windhorst}%%% Rogier.Windhorst@gmail.com
\affiliation{School of Earth and Space Exploration, Arizona State University, Tempe, AZ 85287-1404, USA}

\author{Rui Zhou}
\affiliation{School of Physics and Astronomy, University of Minnesota, 116 Church Street SE, Minneapolis, MN 55455, USA}

\author[0000-0001-5984-0395]{Marusa Bradac}
\affiliation{University of Ljubljana, Department of Mathematics and Physics, Jadranska ulica 19, SI-1000 Ljubljana, Slovenia}
\affiliation{Department of Physics and Astronomy, University of California Davis, 1 Shields Avenue, Davis, CA 95616, USA}

\author[0000-0003-2680-005X]{Gabriel Brammer}
\affiliation{Cosmic Dawn Center (DAWN), Denmark}
\affiliation{Niels Bohr Institute, University of Copenhagen, Jagtvej 128, DK-2200 Copenhagen N, Denmark}

\author[0000-0002-6338-7295]{Victoria Strait}
\affiliation{Cosmic Dawn Center (DAWN), Denmark}
\affiliation{Niels Bohr Institute, University of Copenhagen, Jagtvej 128, DK-2200 Copenhagen N, Denmark}

\author[0000-0002-8785-8979]{Tom J. Broadhurst}
\affiliation{University of the Basque Country UPV/EHU, Department of Theoretical Physics, Bilbao, E-48080, Spain}
\affiliation{DIPC, Basque Country UPV/EHU, San Sebastian, E-48080, Spain}
\affiliation{Ikerbasque, Basque Foundation for Science, Bilbao, E-48011, Spain}

\author[0000-0001-9065-3926]{Jose M. Diego}
\affiliation{Instituto de F\'i sica de Cantabria (CSIC-UC). Avda Los Castros s/n. 39005 Santander, Spain}

\author[0000-0003-1625-8009]{Brenda L.~Frye}
\affiliation{Department of Astronomy/Steward Observatory, University of Arizona, 933 N. Cherry Avenue, Tucson, AZ 85721, USA}

\author[0000-0002-7876-4321]{Ashish K. Meena}
\affiliation{Physics Department, Ben-Gurion University of the Negev, P.O. Box 653, Be'er-Sheva 84105, Israel}

\author[0000-0002-0350-4488]{Adi Zitrin}
\affiliation{Physics Department, Ben-Gurion University of the Negev, P.O. Box 653, Be'er-Sheva 84105, Israel}

\author[0000-0002-2282-8795]{Massimo Pascale}
\affiliation{Department of Astronomy, University of California, 501 Campbell Hall \#3411, Berkeley, CA 94720, USA}

\author[0000-0001-9875-8263]{Marco Castellano}
\affiliation{INAF - Osservatorio Astronomico di Roma, via di Frascati 33, 00078 Monte Porzio Catone, Italy}

\author[0000-0001-9002-3502]{Danilo Marchesini}
\affiliation{Department of Physics and Astronomy, Tufts University, 574 Boston Ave., Medford, MA 02155, USA}

\author[0000-0002-8512-1404]{Takahiro Morishita}
\affiliation{IPAC, California Institute of Technology, MC 314-6, 1200 E. California Boulevard, Pasadena, CA 91125, USA}

\author[0000-0002-8434-880X]{Lilan Yang}
\affiliation{Kavli Institute for the Physics and Mathematics of the Universe, The University of Tokyo, Kashiwa, Japan 277-8583}

%\collaboration{20}{(GLASS-JWST)}



\begin{abstract}
We report the discovery of an extremely magnified star at redshift $z=2.65$ in {\it James Webb Space Telescope (JWST)} NIRISS pre-imaging of the Abell 2744 galaxy-cluster field.  The star's background host galaxy lies on a fold caustic of the foreground lens, and the cluster creates a pair of images of the region close to the lensed star. We identified the bright transient in one of the merging images at a distance of $\sim 0.15''$ from the critical curve, by subtracting the {\it JWST} F115W and F150W imaging from coadditions of archival {\it Hubble Space Telescope (HST)} F105W and F125W images and F140W and F160W images, respectively.
Since the time delay between the two images should be only hours, the transient must be the microlensing event of an individual star, as opposed to a luminous stellar explosion which would persist for days to months. Analysis of individual exposures suggests that the star's magnification is not changing rapidly during the observations. From photometry of the point source through the F115W, F150W, and F200W filters, we identify a strong Balmer break, and modeling allows us to constrain the star's temperature to be approximately 7,000--12,000\,K. 
% ApJL 250 words
\end{abstract}



\section{Introduction} \label{sec:intro}

Galaxies and galaxy clusters act as massive gravitational lenses that are able to magnify intrinsically faint background sources.  Their magnifying power becomes greatest for intrinsically compact sources adjacent to their critical curves (or caustics in the source plane), because regions of greatest magnification are small. Indeed, the possibility that stars that could become highly magnified by galaxy-cluster lenses was suggested in the early 1990's \citep{miraldaescude91}.
%Some of the faint sources at cosmological distances such as single stars can be extremely magnified, when the faint sources and microlenses are aligned and adjacent to the critical curve of a cluster lens. 
%When these faint sources are adjacent to the caustic of a cluster lens, or their images to the cluster's critical curve, and when aligning with a microlens in the cluster, they can become extremely magnified. 
%Thus, the extreme gravitational lensing by galaxy clusters makes them a powerful tool to detect and study individual stars at cosmological distances.

The first example of an extremely magnified star was a blue supergiant in a spiral galaxy at $z=1.49$ (dubbed Icarus) discovered by \citet{kellydiegorodney18}. The star was identified in the {\it Hubble} Space Telescope ({\it HST}) imaging of the MACS J1149.5+2223 galaxy cluster field taken to follow up Supernova (SN) Refsdal \citep{kellyrodneytreu15,treubrammerdiego16,rodneystrolgerkelly16, kellybrammerselsing16}. Microlensing of the background star by a foreground object in the lens caused its magnification to increase by a factor of approximately three to $\sim$2000. The  cluster lens has the effect of boosting the effective lensing effect of foreground microlenses \citep{diegokaiserbroadhurst18,venumadhavdaimiraldaescude17}.
The Spock events in an arc at $z=1$ in the MACS J0416.1-2403 galaxy-cluster fields were probable microlensing events \citep{rodneybalestrabradac18}, and a second star dubbed Warhol ($z=0.94$, \citealp{chenkellydiego19, kaurovdaivenumadhav19}) was identified from microlensing events found in {\it HST} imaging of the galaxy cluster. 
More recently, \citet{2022Natur.603..815W} reported a highly magnified star dubbed Earendel at $z=6.2$ behind galaxy cluster field WHL0137–08. %($z=0.566$)
Unlike the previous examples, microlensing, which provides direct evidence that the source has a size of less than tens of AU, has not yet been detected for Earendel, and the constraints on its size come from the galaxy-cluster lens model. 

With a photon collection area a factor of six greater than that of {\it HST} and sensitivity across the infrared, the {\it James Webb Space Telescope} ({\it JWST}) improves our ability to detect transient events within highly magnified regions of galaxy cluster fields. Through observations of such extreme-magnification events, we can probe individual stars at cosmological distances, including potentially individual Population III stars \citep{windhorsttimmeswyithe18}.

Here we report the discovery of an extremely magnified star at $z=2.65$ in the first set of {\it JWST} images of the Abell 2744 galaxy-cluster field, which were acquired by the GLASS-JWST program (PI: Treu; ERS-1324; \citealp{treurobertsborsanibradac22}). As shown in Fig.~\ref{fig:mosaic}, the event was detected in {\it JWST} NIRISS \citep[the Near Infrared Imager and Slitless Spectrograph][]{NIRISS} imaging at a separation of $\sim 0.15''$ from the inferred location of the critical curve. In this paper,
we describe the {\it JWST} NIRISS pre-imaging in Section~\ref{sec:data}. Section~\ref{sec:analysis} provides the details of our data analysis and results. Our conclusions and discussions are presented in Section~\ref{sec:discussion}.
We assume a concordance $\Lambda$ cold dark matter cosmological model with $\Omega_m=0.3$, $\Omega_\Lambda=0.7$, and a Hubble constant H$_0=70\,\mathrm{km}\,\mathrm{s}^{-1}\,\mathrm{Mpc}^{-1}$. All magnitudes are in the AB system \citep{okegunn83}.

%The frequency of bright microlensing events including Icarus \citep{kellydiegorodney18}, likely the Spock events \citep{rodneybalestrabradac18}, and Warhol provide a new probe of the mass density of objects in the intracluster medium \citep{diegokaiserbroadhurst18,kellydiegorodney18,venumadhavdaimiraldaescude17,oguridiegokaiser18}, as well as the qualitative properties and luminosity functions of massive stars at high redshift \citep{kellydiegorodney18}. 
%\citet{diego18} have found that $\sim$50,000 luminous stars at redshifts between $z=1.5$ and $z=2.5$ should experience an average magnification exceeding 100 from lensing halos of all masses. Of these, approximately 8000 stars should have a mean magnification greater than 250 and should exhibit relatively frequent microlensing peaks. \citet{windhorsttimmeswyithe18} have also recently shown that high magnification during caustic-crossing events close to cluster critical curves should provide an opportunity to observe directly Population III stars at high redshifts using the {\it JWST}.

\section{Data}
\label{sec:data}
The {\it JWST} Director’s Discretionary Early Release Science Program ERS 1324 (PI T. Treu; Through the Looking GLASS: A {\it JWST} Exploration of Galaxy Formation and Evolution from Cosmic Dawn to Present Day) acquired the NIRISS imaging of the Abell 2744 galaxy-cluster field, as the pre-imaging component to slitless spectroscopy \citep{2022arXiv220711387R}[paper I]. A total of 24 NIRISS imaging exposures were acquired from 2022-06-28 22:04:38.674 UTC to 2022-06-29 10:41:13.507 UTC in the F115W, F150W, and F200W filters. Details on observing strategy can be found in \citet{treurobertsborsanibradac22}. We retrieved stage two images and combined the imaging using the {\it JWST} pipeline \footnote{\url{https://jwst-pipeline.readthedocs. io/en/latest/jwst/introduction.html}}. 

\section{Data Analysis and Results}
\label{sec:analysis}

\subsection{A Transient}

\begin{figure*}
\centering
\includegraphics[angle=0,width=6.8in]{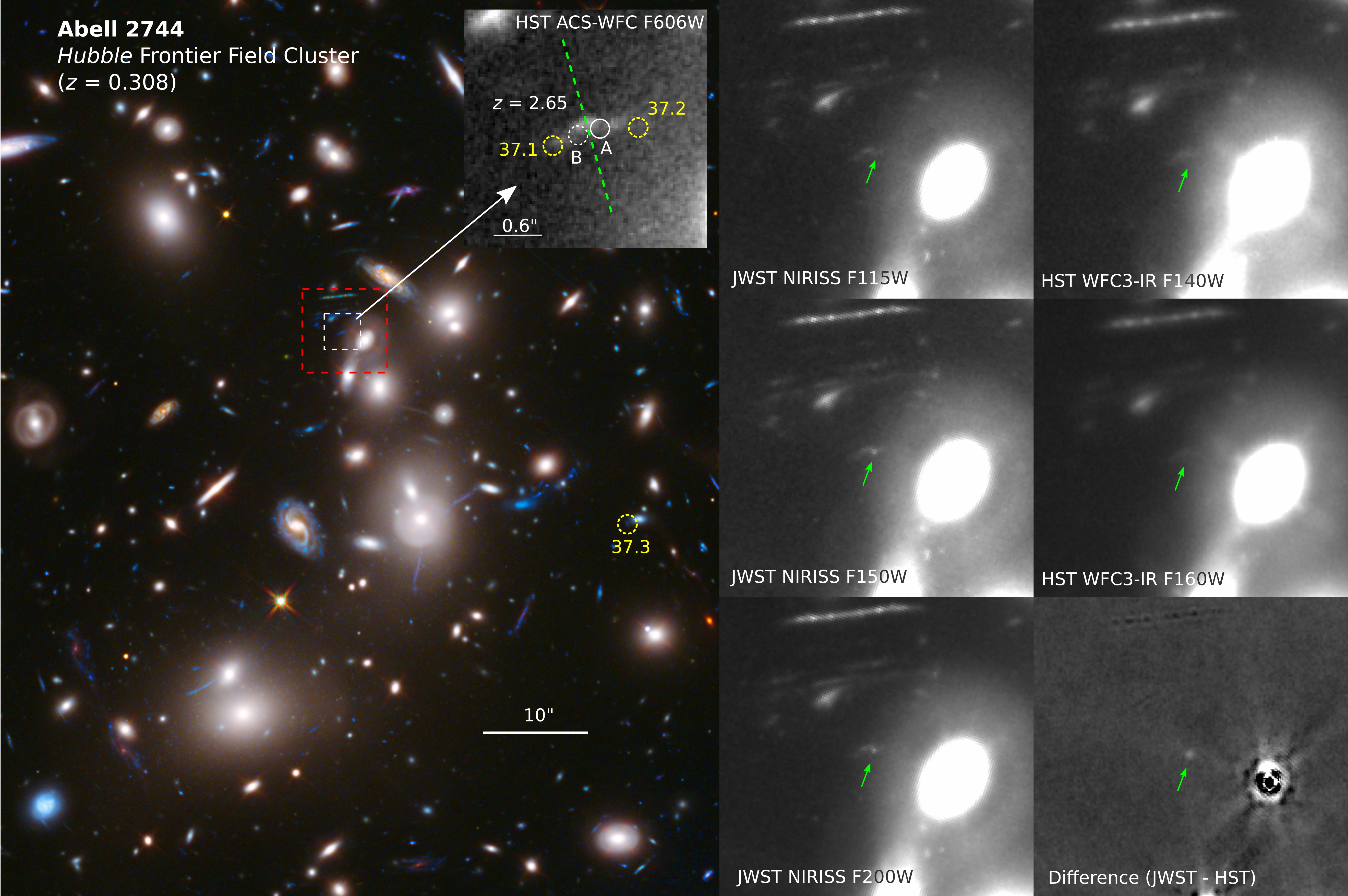}
\caption{Left panel shows the color composite image of the galaxy-cluster Abell 2744 created from the {\it HST} ACS WFC and WFC3 IR cameras and the location of the newly discovered lensed star in an arc at $z=2.65$ \citep{2018MNRAS.473..663M} in the {\it JWST} NIRISS imaging. The six panels on the right show the {\it HST} and {\it JWST} images of the region enclosed by the red dashed lines in the left panel. The three panels in the middle column show the {\it JWST} NIRISS images of the transient (green arrows) acquired by the {\it JWST} ERS-1324 program. The top-right and the middle-right panels are coadded {\it HST} WFC3-IR F140W and WFC3-IR F160W images of the region, respectively. The bottom right panel shows the difference image between the convolved {\it JWST} NIRISS F150W image and the combined {\it HST} template generated from WFC3-IR F140W and WFC3-IR F160W images. The inset in the top right corner of the left panel shows the {\it HST} ACS-WFC F606W imaging of the sky region enclosed by the white dashed lines. Circle $A$ marks the position of the newly detected transient. The green dashed line illustrates the critical curve whose location is inferred from the mirrored images of the arc. Circle $B$ marks the position of the expected counter image of $A$, and we detect no transient at that location. Yellow circles labeled by 37.1, 37.2 and 37.3 are a set of multiple images listed in Tabel A1 in \citet{2018MNRAS.473..663M}.
\label{fig:mosaic}}
\end{figure*}

As shown in Fig.~\ref{fig:mosaic}, the transient we report here was detected in the {\it JWST} NIRISS F150W and F200W imaging. The transient's flux did not vary significantly among the 16 F150W and F200W exposures acquired during the $\sim 0.35$-day visit to the field. There is no statistically significant detection of this transient in F115W imaging.  The transient's coordinates are $\alpha = 0^{\rm h}14^{\rm m}21.326^{\rm s}$, $\delta = -30^{\circ}23'41.46''$ in the World Coordinate System (WCS) of the official images produced by the {\it Hubble} Frontier Field survey collaboration \citep{lotzkoekemoercoe17}.

The transient was identified in the difference image between a convolved {\it JWST} image and the {\it HST} template. To compare {\it HST} images with the {\it JWST} NIRISS F150W image, we combined the {\it HST} WFC3-IR F140W and WFC3-IR F160W coadded images to create a template, since the wavelength range of the {\it JWST} NIRISS F150W filter is spanned by the two {\it HST} WFC3-IR filters. The effective wavelengths of the {\it JWST} NIRISS F150W, {\it HST} WFC3-IR F140W, and {\it HST} WFC3-IR F150W filters are 14846\,\AA, 13734\,\AA, and 15278\,\AA, respectively. We then compute a transition kernel $T$, for which $\mathrm{PSF}_\mathrm{JWST}* T \approx \mathrm{PSF}_\mathrm{HST}$, where $\mathrm{PSF}_\mathrm{HST}$ and $\mathrm{PSF}_\mathrm{JWST}$ are point-spread functions of the two telescopes. The kernel is determined using the Richardson–Lucy algorithm \citep{1972JOSA...62...55R,1974AJ.....79..745L} based on a set of manually selected bright (but not saturated), isolated {\revise sources} present in the imaging from both telescopes. {\revise Details are described in Appendix A.} The transient is visibly apparent (with $\sim 8\sigma$ significance) in the resulting difference image, as shown in the bottom-right panel in Fig.~\ref{fig:mosaic}.

After correcting for Galactic dust extinction ($A_V = 0.0354$\,mag; \citealt{schlaflyfinkbeinerSFD11}), we obtain flux densities within a 0.12$''$ aperture centered on the transient position of $8.98\pm7.58$ nJy, $55.09\pm7.13$ nJy, and $57.28\pm7.66$ nJy for the NIRISS F115W, F150W, F200W filters, respectively.

\subsection{Underlying Arc}

{\revise The redshift of the underlying arc was measured to be 2.6501 from spectroscopy acquired using the Low Resolution Imaging Spectrometer (LRIS) mounted on the Keck-I telescope} \citep{2018MNRAS.473..663M}. % during the night of 2015 December 7
A so-called fold caustic creates mirror, merging images of a galaxy that intersects it. Such a pair of mirrored images allow us to estimate the position of the critical curve. For the arc where this event has been detected, we are able to identify the critical curve's location based on the arc's morphology, as shown by the green dashed line in the inserted panel in Fig.~\ref{fig:mosaic}, where $A$ and $B$ label the positions of the transient and its mirrored image on opposite sides of the critical curve, respectively. In the {\it JWST} observation, no counter image of the transient is detected.

We measure the photometry of the arc based on the official coadded images taken from the {\it Hubble} Frontier Field survey \citep{lotzkoekemoercoe17}. We find the total stellar mass and the star-formation rate of the arc are $7.4\times10^7 M_\odot$ and $4.5 M_\odot\cdot {\rm yr}^{-1}$, respectively, without correction for lensing magnification. The surface stellar-mass density of the cluster at the position of the arc is $1.1\times10^7 M_\odot\cdot {\rm kpc}^{-2}$.

\subsection{Gravitational Lens Models}

\begin{figure*}
\centering
\includegraphics[angle=0,width=6.8in]{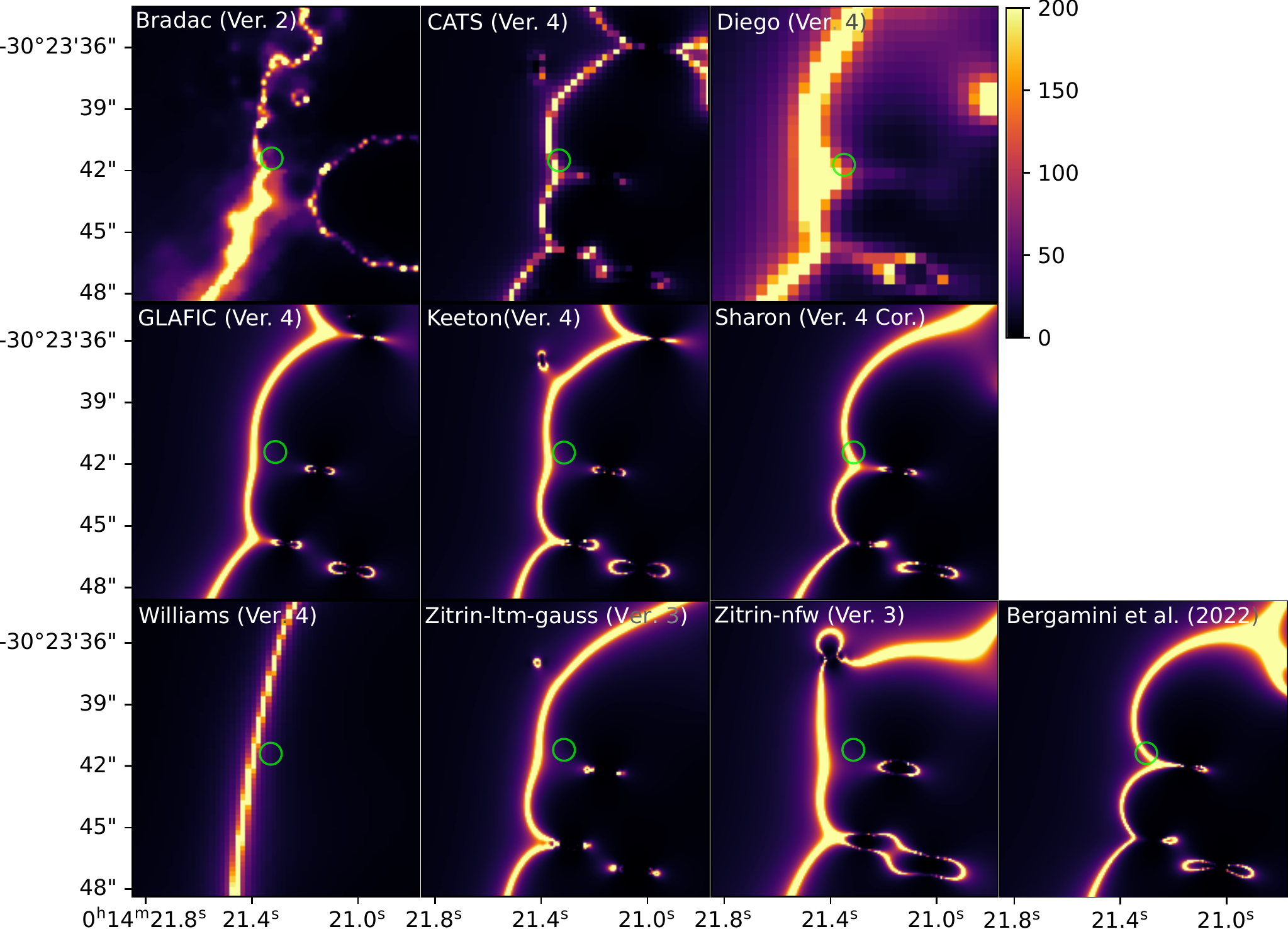}
\caption{Magnification maps for published lens models in the vicinity of the transient. The newly discovered transient event in the {\it JWST} NIRISS imaging marked by the green circles is close to the critical curve of the galaxy cluster Abell 2744.
\label{fig:mag_maps}}
\end{figure*}

Using the published Frontier Fields Lens Models\footnote{\url{https://archive.stsci.edu/pub/hlsp/frontier}} \citep{lotzkoekemoercoe17} listed in Table~\ref{tab:mag}  and a recent model \citep{Bergamini22} for the Abell 2744 galaxy cluster, we calculate their magnification maps at $z=2.65$ and compare these maps with the position of the transient, as shown in Fig.~\ref{fig:mag_maps}. The predicted magnification $\mu$ at the position of the newly discovered transient is listed in Table~\ref{tab:mag}. For newly released model from \citet{Bergamini22}, we obtained the magnification the best-fit model generated from the paper's Strong Lensing Online Tool. We can see that the position of the event is close to the predicted galaxy-cluster critical curves from these lens models. The model predictions are also, in general, consistent with the multiple images of the arc, although these models' predictions for the position of the critical curve have large uncertainties. We also note that the Frontier Fields Lens Models were constructed before the spectroscopic redshift of this system was known.

The position of the transient, marked by $A$ in the inserted image in the left panel of Fig.~\ref{fig:mosaic}, is $\sim 0.15''$ from the inferred location of the critical curve as shown by the green dashed line. We evaluate the magnification at the position offset by $0.15''$ from these models' predicted critical curve. As listed in Table~\ref{tab:mag}, except Diego V4 and Zitrin-nfw V3, all the other models predict a magnification of $\sim 100-200$ at $0.15''$ from the critical curve, while Diego V4 and Zitrin-nfw V3 favor even greater magnifications.

Furthermore, for models with a resolution smaller than $0.1''\,\mathrm{pixel}^{-1}$, we compute the relative time delay between position $A$ and its counter-image $B$ as labeled in the inserted image in the left panel of Fig.~\ref{fig:mosaic}. We find that all the selected models predict a relative delay of less than 4 hours. This is smaller than the duration of the {\it JWST} observation. Thus, if the event were a supernova in the arc, we should detect a counter-image at the $B$ position in the {\it JWST} imaging. The non-detection of the counter-image of the transient at the other side of the critical curve allows us to exclude a multiply imaged stellar explosion.

As shown in the left panel in Fig.~\ref{fig:mosaic}, the arc is consistent with a set of multiple images (37.1, 37.2, and 37.3). According to Table A1 in \citet{2018MNRAS.473..663M}, 37.1 and 37.2 have the same spectroscopic redshift $z=2.6501$, while the redshift of 37.3 has not yet been confirmed.
%As shown in the left panel in Fig.~\ref{fig:mosaic}, from \citet{2018MNRAS.473..663M}, there are a set of multiple images 37.1, 37.2, and 37.3, where 37.1 and 37.2 have confirmed spectroscopic redshift of $z=2.65$. Positions of 37.1 and 37.2 are, in general, consistent with the critical curve we obtained from the arc's morphology.
%However, the spectroscopic redshift of the galaxy at the position of 37.3 according to \citet{2015ApJ...811...29W} is 0.508 \citep{2018MNRAS.473..663M}, which cannot be the counter-image of the arc.

\begin{deluxetable*}{lcccccc}
\tablecaption{{\revise Gravitational-lensing properties from published lens models:} Magnification ($\mu$), the 68\% confidence interval (CI) of magnification, {\revise convergence ($\kappa$), and shear ($\gamma$)} at the location of the transient; magnification at 0.15\arcsec\ from the local critical curve ($\mu'$), and relative time delay ($dt$) between a pair of counter images at 0.15\arcsec\ from the local critical curve.\tablenotemark{a}. \label{tab:mag}}
\tablecolumns{7}
\tablewidth{0pt}
\tablehead{
\colhead{Model} &
\colhead{$\mu$} &
\colhead{68\% CI of $\mu$} &
\colhead{$\kappa$} &
\colhead{$\gamma$} &
\colhead{$\mu'$} &
\colhead{$dt$ (hours)}
}
\startdata
\multicolumn{7}{c}{Before MUSE Spectroscopy}\\
Brada{\v c} (v2)\tablenotemark{b} & 55 & {\revise (35, 188)} & {\revise 0.76} & {\revise 0.28} & 172 & N/A \\
Zitrin-ltm-gauss (v3)\tablenotemark{c} & 47 & (36, 204) & {\revise 0.73} & {\revise 0.34} & 195 & 1.68 \\
Zitrin-nfw (v3)\tablenotemark{d} & 19 & (14, 26) & {\revise 0.85} & {\revise 0.24} & 270 & 3.14 \\
\hline
\multicolumn{7}{c}{After MUSE Spectroscopy}\\
CATS (v4)\tablenotemark{e} & 64 & (50, 89) & {\revise 0.79} & {\revise 0.26} & 110 & N/A \\
Diego (v4)\tablenotemark{f} & 84 & (75, 93) & {\revise 0.89} & {\revise 0.16} & 710 & N/A \\
GLAFIC (v4)\tablenotemark{g} & 22 & (20, 24) & {\revise 0.80} & {\revise 0.28} & 184 & 0.59 \\
Keeton (v4)\tablenotemark{h} & 31 & (23, 47) & {\revise 0.80} & {\revise 0.27} & 178 & 0.56 \\
Sharon (v4 Cor.)\tablenotemark{i} & 282 & (172, 730) & {\revise 0.73} & {\revise 0.29} & 161 & 2.95 \\
Williams/GRALE (v4)\tablenotemark{j} & 26 & (18, 86) & {\revise 0.83} & {\revise 0.26} & 160 & N/A \\
{\revise Bergamini et al. (2020)\tablenotemark{k}} & {\revise 122} & {\revise (70, 423)} & {\revise N/A} & {\revise N/A} & {\revise N/A} & N/A \\
\enddata
\tablenotetext{a}{{\revise $\mu$ is the median value, while $\kappa$, $\gamma$,} $\mu'$, and $dt$ are based on the best-fit models available from \url{https://archive.stsci.edu/pub/hlsp/frontier/abell2744/models/}. We only evaluate the relative time delay of the Frontier Field Lens Models \citep{lotzkoekemoercoe17} with resolution smaller than $0.1\arcsec \mathrm{pixel}^{-1}$. The horizontal line separates models before and after the Multi Unit Spectroscopic Explorer {\revise (MUSE)} observations of {\revise Abell 2744} \citep{2018MNRAS.473..663M}.}
\tablenotetext{b}{\cite{Wang.2015mhw,hoaghuangtreu16,bradactreuapplegate09,bradacschneiderlombardi05}}
\tablenotetext{c}{\cite{zitrinmeneghettiumetsu13,zitrinbroadhurstumetsu09} (see also \citealt{mertencoedupke11,mertencacciatomeneghetti09})}
\tablenotetext{d}{\cite{zitrinmeneghettiumetsu13,zitrinbroadhurstumetsu09} (see also \citealt{mertencoedupke11,mertencacciatomeneghetti09})}
\tablenotetext{e}{\cite{2018MNRAS.473..663M}}
\tablenotetext{f}{\cite{diegoprotopapassandvik05,diegosandvikprotopapas05,diegotegmarkprotopapas07,diegobroadhurstbenitez15}}
\tablenotetext{g}{\cite{kawamataishigakishimasaku18,kawamataoguriishigaki16,oguri10}}
\tablenotetext{h}{\cite{mccullykeetonwong14,ammonswongzabludoff14,keeton10}}
\tablenotetext{i}{\cite{johnsonsharonbayliss14,jullokneiblimousin07}}
\tablenotetext{j}{\cite{sebesta16,liesenborgsderijckedejonghe06}}
{\revise \tablenotetext{k}{\cite{Bergamini22}}}
\end{deluxetable*}

\subsection{Spectral Energy Distribution of the Lensed Star}

\begin{figure*}
\centering
\includegraphics[angle=0,width=3.5in]{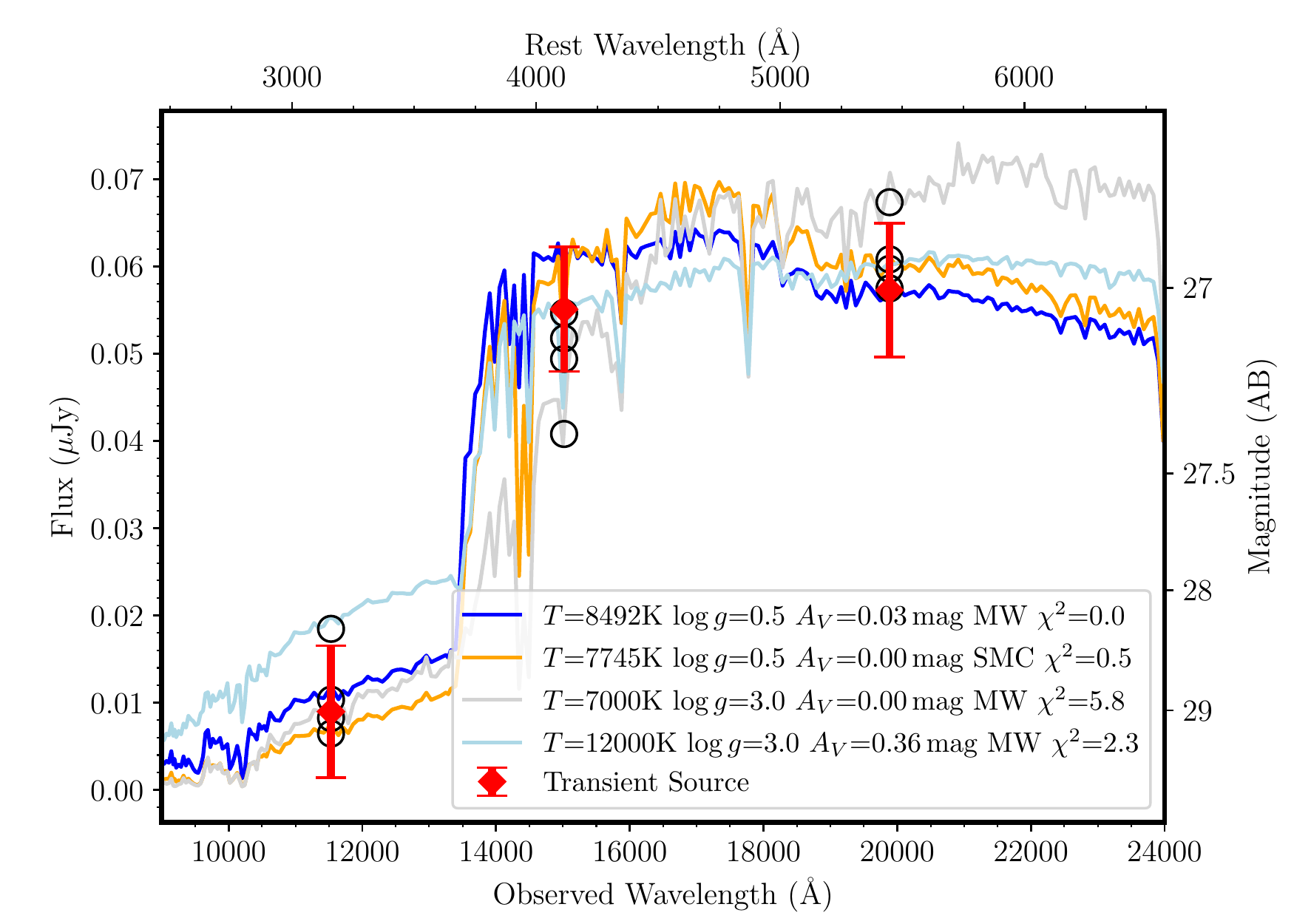}
\includegraphics[angle=0,width=3.5in]{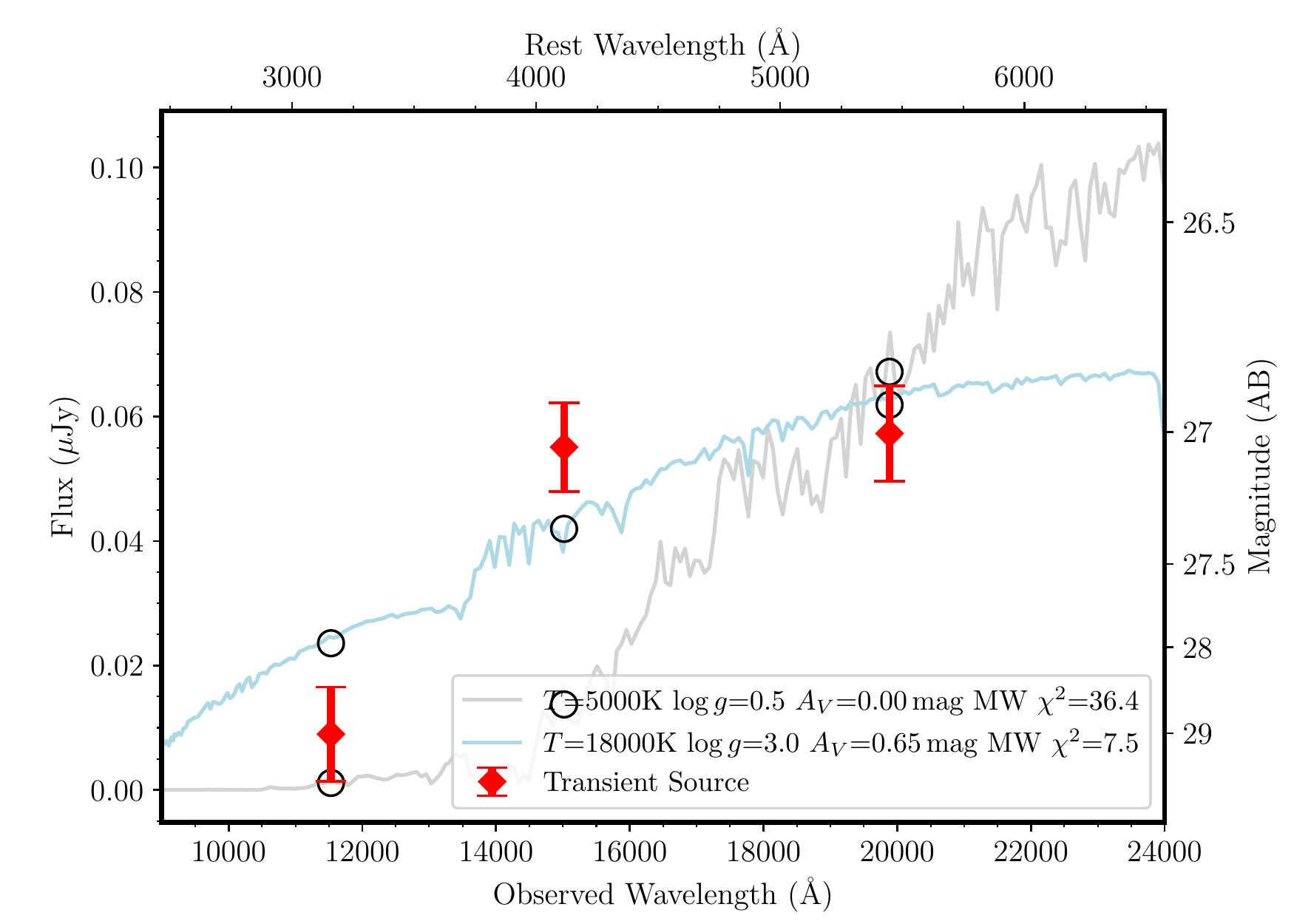}
\caption{Stellar atmosphere models and the measured SED of the transient. Left panel shows 7000\,K and 12000\,K models with an $R(V)=3.1$ extinction law, as well as best-fitting 8,500\,K and 7,700\,K temperatures for extinction laws with $R(V)=3.1$ and $R(V)=2.73$, respectively. The 7,000\,K and 12,000\,K models are consistent with the measurements within $\sim$1--2$\sigma$. The red data points are the photometry of the transient measured from the {\it JWST} NIRISS F115W, F150W, and F200W imaging. Panel at right shows models with 5,000\,K and 18,000\,K where we allow the extinction to vary, and these models are in greater tension with the measurements.
\label{fig:sed}}
\end{figure*}

We fit the spectral energy distribution (SED) of the lensed-star candidate measured from the {\it JWST} NIRISS F115W, F150W, and F200W imaging. A stellar atmosphere model \citep{lejeunecuisinierbuser98} and a host-galaxy extinction curve are simultaneously fit into the measured SED of the lensed star.
As shown in Fig.~\ref{fig:sed}, we find the best-fit photospheric temperature of 8,500\,K for a MW ($R(V)=3.1$; \citealt{cardelli89}) extinction law, and the best-fit photospheric temperature of 7,700\,K for a SMC ($R(V)=2.7$; \citealt{gordonclaytonmisselt03}) extinction law). Fig.~\ref{fig:sed} also shows that models with 7,000\,K and 12,000\,K temperatures are also consistent with 1--2\,$\sigma$ with the measured fluxes. 

\subsection{Constraints on the source size and magnification}

The flux of the transient in the F150W and F200W imaging did not vary significant across exposures, and hence the total time duration of the event is unconstrained. However, we can determine a lower limit on the event duration. The transient has been detected across a period of $\sim 0.35$ days (8.4 hours) from the first F150W exposure to the last F200W exposure. Thus, the duration of the microlensing peak must be $>0.35$ days.

At an offset of $0.15''$ from the galaxy-cluster critical curve near the position of the transient, we obtain $\mu \approx 200$ ($\mu_{t} \approx 100$, $\mu_r \approx 2$, according to the GLAFIC V4 model, where $\mu_t^{-1}=1-\kappa-\gamma$ and $\mu_r^{-1}=1-\kappa+\gamma$, and $\kappa$ and $\gamma$ are convergence and shear), where the Einstein radius of a microlens should be exaggerated by the cluster lens, and a microlensing peak will appear during the caustic crossing, as described in \citet{oguridiegokaiser18,diegokaiserbroadhurst18,venumadhavdaimiraldaescude17,2022MNRAS.514.2545M}.

{\revise \citet{oguridiegokaiser18} derived a pair of equations that enable constraints on the size of the lensed source, given limits on its duration, and its maximum magnification. However, these equations only apply when the microlensing arises from a single microlens (i.e., the microcaustics are not overlapping). We measure a stellar-mass density of $11.55^{+7.61}_{-8.65}\,M_\odot\,\mathrm{pc}^{-2}$ and infer an optical depth for microlensing of $0.52^{+0.34}_{-0.39}$ at the position of the transient (as described in detail in Appendix B). Consequently, we can have a reasonable expectation that the simplified, \citet{oguridiegokaiser18} analysis should yield accurate inferences. We note that \citet{diegokaiserbroadhurst18} and \citet{venumadhavdaimiraldaescude17} provide full treatments of microlensing through simulations.}

The source crossing time for the caustic-crossing event, according to \citet{oguridiegokaiser18}, is given by
\begin{equation}
t_{\rm src} \approx 0.024 \left(\frac{R_{\rm source}}{R_{\odot}}\right) \left(\frac{v}{500\,{\rm km}\, {\rm s}^{-1}}\right)^{-1}\,{\rm days}.
\end{equation}
Thus, the minimum radius of the source is $14\,R_\odot\times(v / 500\,{\rm km}\, {\rm s}^{-1})$ given the minimum 0.35-day event duration and the assumption that the characteristic velocity is 500$\,{\rm km}\, {\rm s}^{-1}$.
Abell 2744 is a complex galaxy cluster merger with a characteristic velocity of $\sim 4000\,{\rm km}\, {\rm s}^{-1}$ \citep{2011MNRAS.417..333M} as inferred from measurements of the galaxies' velocities
and that of the gas in the intracluster medium. The southern ``clump,'' where we detect the transient, has a peculiar velocity of $\sim 2000\,{\rm km}\, {\rm s}^{-1}$ as reported by \citet{2011ApJ...728...27O}. Assuming the cluster's transverse velocity of $v = 2000\,{\rm km}\, {\rm s}^{-1}$, the minimum radius of the source is $56\,R_\odot$.
Moreover, from Eq.\,26 in \citet{oguridiegokaiser18}, the maximum magnification during a caustic crossing for the lensing geometry of this event is given by
\begin{equation}
\mu_{\rm max} \approx 6.1 \times 10^4 \left(\frac{M_{\rm lens}}{M_{\odot}}\right)^{1/4}  \left(\frac{R_{\rm source}}{R_{\odot}}\right)^{-1/2},
\label{eq:mu_max}
\end{equation}
where $M_{\rm lens}$ is the mass of the micro lens.
We note that the arc is behind the outskirt of a bright galaxy in the cluster. Assuming a micro lens with $M_{\rm lens} = 1\,M_\odot$, we have the maximum magnification $\mu_{\rm max} \approx 20,000$, $6,000$, and $2,000$ for $R_{\rm source} = 10\,R_{\odot}$, $100\,R_{\odot}$, and $1,000\,R_{\odot}$, respectively. For $R_{\rm source} = 56\,R_{\odot}$, we have the maximum magnification $\mu_{\rm max} \approx 8,000$. For a more massive microlens with $M_{\rm lens} \gtrsim 3\,M_\odot$, we have the maximum magnification $\mu_{\rm max} \gtrsim 10,000$.

In addition, for a solar-mass microlens and a $2000\,{\rm km}\, {\rm s}^{-1}$ transverse velocity of the cluster lens, the time between caustic crossings \citep{oguridiegokaiser18} will be as long as $\sim 300\,{\rm days}$. 

\subsection{Luminosity and Magnification of Lensed Star}

For stars of different spectral types and absolute magnitudes, we evaluate the required magnification to reach the observed apparent magnitude of $27.05$ in the {\it JWST} NIRISS F150W imaging. We perform the same calculation as that in \citet{chenkellydiego19}, and our results are listed in Table~\ref{tab:stars}.
%To determine the star's magnification, we first calculate a {\it K}-correction $K_{\rm xy}$ as 
%\begin{equation}
%m_{\rm y} = M_{\rm x} + dm + K_{\rm xy},
%\end{equation}
%where $m_{\rm y}$ is the observer-frame apparent magnitude in the {\rm y} band, $M_{\rm x}$ is the rest-frame absolute magnitude in the {\rm x} band, and $dm$ is the distance modulus. We use Eq. 2 of \citet{kimgoobarperlmutter96},
%\begin{equation}
%K=2.5\,\log_{10}(1+z) + m_{{\rm y,syn}}^{\rm AB} - m_{V, {\rm syn}}^{\rm Vega},
%\end{equation}  
%to calculate the {\it K}-correction, where $z=2.65$, $m_{{\rm y,syn}}^{\rm AB}$ is the synthetic magnitude of a redshifted model spectrum in an observer-frame y band (e.g., {\it JWST} NIRISS F150W), and $m_{V,{\rm syn}}^{\rm Vega}$ is the synthetic Johnson {\it V}-band magnitude of the rest-frame model spectrum. Our results are listed in Table~\ref{tab:stars}.
As we can see, based on the constraints on the minimum source size and the maximum magnification we described above, for a microlens with a few solar masses, a main sequence star is not sufficiently luminous. A very luminous ($M_V \approx -8.5$) blue supergiant (A0 or F5) is required to explain the microlensing event that we have discovered. We note that, at lower metallicities, stars of equal temperature become comparatively more luminous.

\begin{deluxetable*}{ccccccc}
\tablecaption{Magnification ($\mu$) required for different types of stars. Approximate peak magnifications are for no host-galaxy extinction and for the flux density observed F150W magnitude of $\sim27.05$. Consequently, we favor a post-main-sequence blue/blue-white supergiant having $-9 \lesssim M_V \lesssim -7$. \label{tab:stars}}
\tablecolumns{6}
%\tablenum{2}
\tablewidth{0pt}
\tablehead{
&
\colhead{Spec. Model} &
\colhead{Temp} &
\colhead{$M_V$} & 
\colhead{F150W} & 
\colhead{$K$} & 
\colhead{$\mu$} 
}
\startdata
%Type    Temp    AbsMag  AppMag  App-Obs F125W-F160W     kcor    magnif
%Type    Temp    AbsMag  AppMag  App-Obs F115W-F150W     kcor    magnif
Extreme BSG&A5&8491\,K&-8.50&37.04&-1.20&9894\\
Extreme BSG&F0&7211\,K&-8.50&37.22&-1.03&11665\\
\hline
MS&O5V&39810\,K&-5.40&39.33&-2.01&81613\\
MS&O9V&35481\,K&-4.00&40.80&-1.95&315414\\
MS&B0V&28183\,K&-3.70&41.10&-1.94&416393\\
MS&B1V&22387\,K&-3.20&41.73&-1.81&745490\\
MS&B3V&19054\,K&-2.10&42.92&-1.72&2237247\\
MS&B5-7V&14125\,K&-2.10&43.04&-1.61&2479165\\
MS&B8V&11749\,K&-1.08&44.12&-1.54&6720452\\
MS&A5V&8491\,K&2.40&47.94&-1.20&226668196\\
MS&F0V&7211\,K&3.20&48.92&-1.03&558299929\\
\enddata
\end{deluxetable*}

\section{Discussion}
\label{sec:discussion}

The transient we discovered in the {\it JWST} NIRISS pre-imaging can be interpreted as an extremely magnified (by a factor of $\sim 10,000$), luminous ($-9 \lesssim M_V \lesssim -7$) blue/blue-white supergiant (Deneb-like) that was magnified by the Abell 2744 cluster lens and a solar-mass microlens. From the best-fit SED model, as shown in Fig.~\ref{fig:sed}, the Balmer discontinuity at $\sim 14,000$\AA\ in the observer's frame can explain why the lensed star is only visible through {\it JWST} NIRISS F150W and F200W filter but not detected with statistical significance in the F115W imaging. 

We note that our constraints on the source size and the maximum caustic-crossing magnification are based on a solar-mass microlens. A more massive microlens can provide higher magnification which would allow the lensed star to be comparatively less luminous.
Hence, our constraints on the luminosity and types of the lensed star will be relaxed for a massive microlens. In particular, potential populations of primordial black holes with masses of $\sim 1-100\,M_\odot$ would manifest themselves due to their action as microlenses \citep{diegokaiserbroadhurst18}. The discoveries of this kind of lensed stars in {\it JWST} and {\it HST} observations could be used to place constraints on the abundance of primordial black holes as a promising candidate to compact dark matter.
%In addition, the constraint on the source size will be relaxed if the microlens' transverse velocity is relatively small. In the case of a lower transverse velocity, we expect to have a smaller source size and a higher magnification (according to Eq.~\ref{eq:mu_max}). For a large source size or a low transverse velocity, we could also expect to see a long-duration microlensing event that would still be visible in the {\it JWST} or {\it HST} imaging in future visits.

Since we observe no significant evolution in the transient's brightness, we cannot place an upper bound on the event's duration and the source size. Since most $>15\,M_\odot$ stars are observed in binary systems with a separation of $\lesssim 2000\,R_\odot$ \citep{2014ApJS..215...15S,2012Sci...337..444S}, additional transients could become visible within several weeks if the source is composed of multiple stars. 

A potential alternative to a luminous lensed star is a stellar-mass black hole accreting mass from an asymptotic giant branch (AGB) companion. Such systems could potentially be more common than luminous stars since the amount of time lower-mass stars spend on the AGB stage can greatly exceed the lifetimes of high-mass stars \citep{windhorsttimmeswyithe18}. We note that accretion disks of stellar-mass black holes may be difficult to distinguish from massive stars during microlensing events because of their similar temperatures, sizes, and luminosities. However, the continuum spectrum of an optically thick, steady-state accretion disk can be described by the integrated black-body spectra over the disk’s temperature profile, which should not exhibit a significant break in its SED at $\sim 4,000$\AA\ (in the rest frame) as we observed for this object. Thus, our observation does not favor a lensed stellar-mass black hole accretion disk.

%In near-infrared bands, although the {\it JWST} NIRISS imaging have resolved more point sources in the Abell 2744 field compared to the {\it HST} imaging, those newly resolved sources are smoothed out by the transition kernel when we downgrade the {\it JWST} images for a comparison with the {\it HST} templates. As shown in the bottom-right panel Fig.~\ref{fig:mosaic}, since the transient is $\sim 8\sigma$ significant in the difference image, it cannot be a persistent point source that was newly resolved by the {\it JWST}.

It is possible, in principle, that this new transient could instead originate from the bright galaxy in the cluster. However, it is unlikely to be an SN in the galaxy cluster due to its apparent brightness and the non-detection of its rest-frame $I$-band flux.
It is also possible, in principle, that the transient could be a multiply imaged SN where only one image has become highly magnified (or demagnified) by a microlens or subhalo. However, extreme magnification of a SN should be extremely rare, and the large physical size of a SN photosphere means that substantial demagnification or magnification of each image by a microlens becomes improbable. Follow-up observations could constrain the duration of the microlensing event and the existence of any possible counter images.

Microlensing events with lower magnifications should be much more common than events with greater magnifications. {\it JWST}'s high angular resolution, wavelength coverage in the infrared, and sensitivity should enable the collection of an extensive sample of highly magnified stars.

\begin{acknowledgments}

This work is based on observations made with the NASA/ESA/CSA James Webb Space Telescope. The data were obtained from the Mikulski Archive for Space Telescopes at the Space Telescope Science Institute, which is operated by the Association of Universities for Research in Astronomy, Inc., under NASA contract NAS 5-03127 for JWST. These observations are associated with program JWST-ERS-1324. The specific observations analyzed can be accessed via \dataset[https://doi.org/10.17909/y6dh-6g16]{https://doi.org/10.17909/y6dh-6g16}. We acknowledge financial support from NASA through grant JWST-ERS-1324. Archival images from the Hubble Space Telescope were also used. {\revise We would like to thank Dr. Pietro Bergamini, Prof. Piero Rosati, Prof. Claudio Grillo, Dr. Ana Acebron, and Dr. Eros Vanzella for their helpful comments on our paper and for sharing the predictions of their lens model.}

WC acknowledges support from NASA HST grant AR-15791. PLK is supported by NSF grant AST-1908823 and NASA/Keck JPL RSA 1644110. RAW acknowledges support from NASA JWST Interdisciplinary Scientist grants NAG5-12460, NNX14AN10G and 80NSSC18K0200 from GSFC. JMD acknowledges the support of project PGC2018-101814-B-100 (MCIU/AEI/MINECO/FEDER, UE) Ministerio de Ciencia, Investigaci\'on y Universidades. This project was funded by the Agencia Estatal de Investigaci\'on, Unidad de Excelencia Mar\'ia de Maeztu, ref. MDM-2017-0765. AK is supported by scientist grants NAG5-12460, NNX14AN10G and 80NSSC18K0200 from GSFC. AZ and AKM acknowledge support by Grant No. 2020750 from the United States-Israel Binational Science Foundation (BSF) and Grant No. 2109066 from the United States National Science Foundation (NSF), and by the Ministry of Science \& Technology, Israel. MB acknowledges support from the Slovenian national research agency ARRS through grant N1-0238.
\end{acknowledgments}


\vspace{5mm}
\facilities{{\it HST} (ACS-WFC, WFC3-IR), {\it JWST} (NIRISS imaging)}


\software{jwst 1.6.2 (\url{https://jwst-pipeline.readthedocs. io/en/latest/jwst/introduction.html}), FAST++ 1.3.2 (\url{https://github.com/cschreib/fastpp})}



\appendix

\section{Differencing between the {\it HST} WFC3-IR and {\it JWST} NIRISS imaging}

To search for transients, we convolve and then subtract the {\it JWST} F115W and F150W imaging from coadditions of archival {\it HST} F105W and F125W images and F140W and F160W images, respectively. The {\it JWST} NIRISS imaging has much sharper angular resolution compared to the {\it HST} WFC3-IR imaging. We therefore convolve the {\it JWST} imaging with a transition kernel $T$, for which
\begin{equation}
    \mathrm{PSF}_\mathrm{JWST}* T \approx \mathrm{PSF}_\mathrm{HST},
    \label{trans_kernel}
\end{equation}
where $\mathrm{PSF}_\mathrm{HST}$ and $\mathrm{PSF}_\mathrm{JWST}$ are point-spread functions of the two telescopes, and $*$ denotes the convolution operation. For a source $f$, we have
\begin{equation}
    (f*\mathrm{PSF}_\mathrm{HST}) \approx T*(f*\mathrm{PSF}_\mathrm{JWST}),
    \label{trans_kernel_source}
\end{equation}
where we used the commutative and associative properties of convolution.
In Eq.~\ref{trans_kernel_source}, we note that $f*\mathrm{PSF}_\mathrm{JWST}$ and $f*\mathrm{PSF}_\mathrm{HST}$ are observations of $f$ present in the {\it JWST} and {\it HST} imaging, respectively. Thus, we can utilize sources from the {\it JWST} and {\it HST} imaging to determine the kernel $T$.

For this paper, we determine the kernel using the Richardson–Lucy iteration algorithm \citep{1972JOSA...62...55R,1974AJ.....79..745L}. An observed image can be written as
\begin{equation}
    d = u*P + \delta \approx u*P,
    \label{obsered_im}
\end{equation}
where $d$, $u$, and $P$, and $\delta$ are the observation, source, PSF, and noises, respectively.  
For an iteration number $t$, the estimate of $u$ (denoted by $\hat{u}$) can be written as
\begin{equation}
    \hat{u}^{(t+1)} = \hat{u}^{(t)}\cdot\left(\frac{d}{\hat{u}^{(t)}*P}*P'\right),
    \label{rickardonlucy}
\end{equation}
where $P'$ is the flipped PSF. We applied Eq.~\ref{rickardonlucy} on Eq.~\ref{trans_kernel_source} to solve for $T$ using the iteration, for which $u=T$, $d=f*\mathrm{PSF}_\mathrm{HST}$, and $P=f*\mathrm{PSF}_\mathrm{JWST}$. To minimize the accumulation of noise during the iteration, we manually selected a set of several bright (but not saturated) and isolated sources including stars and galaxies present in the imaging from both telescopes, and then coadded their images for the iteration. %Wenlei: not sure what stacked means here
We note that, in Eq.~\ref{trans_kernel_source}, $f$ need not be a point source, and hence $f*\mathrm{PSF}_\mathrm{JWST}$ and $f*\mathrm{PSF}_\mathrm{HST}$ may be either images of stars or galaxies.

\section{Optical Depth of Microlensing}
The optical depth of microlensing \citep{diegokaiserbroadhurst18} can be given by
\begin{equation}
\tau=\int_0^{D_L}\Omega_En(D_L)dD_L,
\label{ml_optiocal_depth}
\end{equation}
where $\Omega_E$ is the solid angle covered by the Einstein ring from microlenses, and $n(D_L)$ is the number density of the microlenses at an angular-diameter distance $D_L$. Assuming that all mass along the line of sight is concentrated in the cluster lens, Eq. \ref{ml_optiocal_depth} can be simplified as
\begin{equation}
\tau\approx\mu\pi \theta_E^2n_L,
\label{ml_optiocal_depth_simple}
\end{equation}
where $\mu$ is the magnification of the macro lens from the cluster, $\theta_E$ is the Einstein radius, and $n_L$ is the surface number density of the microlenses in the cluster lens. We note that, for point-mass microlenses, $\theta_E^2n_L$ is proportionally to the surface mass density of microlenses.

We measured the photometry in the vicinity of the transient including emissions from the wings of cluster member galaxies and from the intracluster light, and then fit it using the {\tt FAST++} software \citep{2009ApJ...700..221K,fastpp} to obtain the stellar mass. We obtained a surface stellar-mass density of $11.55^{+7.61}_{-8.65}\,M_\odot\,\mathrm{pc}^{-2}$. Plugging it into Eq. \ref{ml_optiocal_depth_simple} (assuming solar-mass microlenses) and for $\mu\sim 10^2$, we have the microlensing optical depth of $0.52^{+0.34}_{-0.39}$ at the position of the transient. 

\bibliography{sample631}{}
\bibliographystyle{aasjournal}



\end{document}